\documentclass[12pt]{article}
\usepackage[pctex32]{graphics}
\begin{document}
\vspace{2cm}
\begin{center}
~
\\
{\bf  \Large Wilson-t'Hooft  Loops in Finite-Temperature Non-commutative Dipole Field Theory from Dual Supergravity}
\vspace{1cm}

                      Wung-Hong Huang\\
                       Department of Physics\\
                       National Cheng Kung University\\
                       Tainan, Taiwan\\

\end{center}
\vspace{1cm}
We first study the temporal Wilson loop in the finite-temperature non-commutative dipole field theory from the  string/gauge correspondence. The associated dual supergravity  background is constructed from the near-horizon geometry of near-extremal D-branes, after applying T-duality and  smeared twist.  We investigate the string configuration therein and find that while the temperature produces a maximum distance $L_{max}$ in the interquark distance the dipole in there could produce a minimum distance $L_{min}$. The quark boundary pair therefore could be found only if their distance is between $L_{min}$ and $L_{max}$.  We also show that, beyond a critical temperature the quark pair becomes totally free due to screening by thermal bath.  We next study the spatial Wilson loop  and find the confining nature in the zero temperature 3D and 4D non-supersymmetry  dipole gauge theory.  The string tension of the linear confinement potential is obtained and found to be a decreasing function of the dipole field.  We also investigate the associated t'Hooft  loop and determine the corresponding  monopole anti-monopole potential.  The conventional screening of magnetic charge which indicates the confinement of the electric charge is replaced by a strong repulsive however.  Finally, we show that the dual string which is rotating along the dipole deformed $S^5$ will behave as a static one without dipole field, which has no minimum distance and has larger energy than a static one with dipole field.  We discuss the phase transition between these string solutions.

\vspace{1cm}
\begin{flushleft}
*E-mail:  whhwung@mail.ncku.edu.tw\\
\end{flushleft}


\newpage
\section{Introduction}
The expectation value of Wilson loop is  one of the most important observations in the gauge theory.   In the AdS/CFT duality [1-3]  it becomes tractable to understand this highly nontrivial quantum field theory effect through
a classical description of the string configuration in the AdS background.  Using the AdS/CFT duality Maldacena [4]  derived for the first time the expectation value of the rectangular Wilson loop operator from the Nambu-Goto action in which the string worldsheet is bounded by a loop and along a geodesic on the $AdS_5 \times S^5$ with endpoints on $AdS_5$.   It is found that the interquark potential exhibits the Coulomb type behavior expected from conformal invariance of the gauge theory.

In order to make contact with Nature many investigations had gone beyond the initial conjectured duality and generalized the method to investigate the theories breaking conformality and (partially)  supersymmetry [5-8].  For example, the Klebanov--Witten solution [9], the Klebanov--Tseytlin solution [10], the Klebanov--Strassler solution [11], and Maldacena--N\'u\~nez (MN) solution [12] which dual to the ${\cal N}=1$ gauge theory.

Historically, the first literature to find a confining theory was discussed by Witten [13] in which  the finite temperature system which break the conformal symmetry of the theory was considered.  The Maldacena's computational technique was then extended to the finite temperature case by replacing the AdS metric by a Schwarzschild-AdS metric [13-17].  Note that in considering the finite temperature theory one direction (the Euclidean time) is compactified along a circular of radius $1/2\pi T$.    Thus, at high temperature the original 4D (5D) theory corresponds essentially to the 3D (4D) theory.   As both the fermions and scalars get a mass of the order of temperature this theory reduces to a pure gauge field theory.  Also, the bosons have periodic and fermions anti-periodic boundary condition, in going to a finite temperature theory we also break the supersymmetry.  Thus, in the high-temperature limit the associated spatial Wilson loop describes the zero-temperature non-supersymmetric gauge theory in 3D (4D) dimension, which then reveals the nature of quark confinement.

The Wilson loop in the  non-commutative dipole field theory from the  string/gauge correspondence had been investigate.   First, the dual supergravity background of the non-commutative dipole theory had been found in [18-20].  Alishahiha and Yavartanoo [20] had also found the general dual supergravity of Dp branes in the presence of a nonzero B field with one leg along the brane worldvolume and other transverse to it, which duals to a non-commutative dipole theory.  They had investigated the associated Wilson loop and found that when the distance between quark and anti-quark is much bigger then their dipole size the energy will show a Coulomb type behavior with a small correction form the non-commutativity.   In the previous papers [21]  we re-examined the problem and found that it exists a minimum distance between the quarks.  We also find that the dual string which is rotating along the dipole deformed $S^5$ will behave as a static one without dipole and has higher energy than the static one with dipole field.  In this paper, we will follow the method in [14-16] to extend our investigation to the finite-temperature non-commutative dipole field theory from the  string/gauge correspondence.  

In section II  we first construct the dual supergravity background of the finite temperature non-commutative dipole theory by considering the near-horizon geometry of near-extremal D-branes, after applying T-duality and  smeared as that described in [18-20].    We study the temporal Wilson loop in the  string/gauge correspondence by investigating the associated string configuration.   We find that while the temperature produces a maximum distance $L_{max}$ the dipole could produce a minimum distance $L_{min}$. The quark boundary pair therefore could be found only if their distance is between $L_{min}$ and $L_{max}$.  Especially, we show that, beyond a critical temperature the quark pair becomes totally free due to screening by thermal bath.  

In section III we study the spatial Wilson loop  and see that the confining nature could be shown in the zero temperature 3D and 4D non-supersymmetry dipole gauge theory.   We obtain the string tension which is found to be a decreasing function of the  dipole field. 

In section IV we follow the method in [16] to  investigate the associated t'Hooft  loop in the 4D non-supersymmetry gauge theory which shows the nature of quark confinement.  The t'Hooft  loop is the ``electric-magnetic" dual of the Wilson loop and describes the monopole anti-monopole potential.  The string theory realized of the monopole  is the D2-brane ending on the D4-brane.  The D2-brane is wrapped along $x_0$ so from the point of view of the 4D theory it is a point like object.   We find that the expectation value of t'Hooft  loop shows strong repulsive force between the monopole and anti-monopole, contrast to the conventional of screening of magnetic charge. 

In section V we study the dual string which is rotating along the dipole deformed $S^5$ and see that it will behave as a static one without dipole field.   We find that it has no minimum distance and has larger energy than a static one with dipole field.  Collecting the above analysis we discuss the phase transition between these string solutions.  Last section is devoted to a summary.

Note that it is a long-belief that in quantum theories including gravity, spacetime must change its nature at distances comparable to the Planck scale. Quantum gravity has an uncertainty principle which prevents one from measuring positions to better accuracies than the Planck length. Thus, the quantum effects could be modeled by a non-commutation relation and non-local properties. String theory is not local and it was discovered in [22,23] that simple limits of M theory and string theory lead directly to noncommutative gauge theories.  As we now known that the noncommutative gauge theories can be realized in string theory as the world volume of D-branes in a constant background B field. It was found by [18] that when the B-field has one leg along the brane and the other transverse to it the noncommutative dipole field theory (NCDFT) appears.  NCDFT's  are also interesting by themselves. It has a chance of finding a CP (and even CPT) violating theory [19]. It is also an appropriate candidate to study the interaction of a neutral particles with finite dipole moments, like neutrinos, with gauge particles like photons. There are some experimental evidences of such interactions, which cannot be described by the commutative version of the standard model of particles [24].

\section {Temporal Wilson Loop in Finite-Temperature Dipole Theory}
\subsection {Supergravity Solution} 
To find the explicit supergravity solution of D3-brane describing the finite temperature dipole theory we could start with the following type II supergravity 
solution describing N coincident near extremal D3-brane  [25] 
$$ds^2= f(r)^{-1/2}\left[- h(r)dt^2+dx_1^2+dx_2^2+dx_3^2\right] + f(r)^{1/2}\left[h(r)^{-1}dr^2+ r^2 d\Omega_5^2\right],$$
$$ f(r) = 1+ {N^{4}\over r^4},~~~~~ h(r) = 1 - {r_0^{4}\over r^4}, \eqno{(2.1)}$$
in which $dr$ and $d\Omega$ constitute $x_4, ..., x_9$ coordinates. The horizon is located at $r =r_0$ and extremality is achieved in the limit $r_0 \rightarrow 0$. A solution with $r_0 \ll N$ is called near extremal.  

Now, as described in [18-20], we first apply the T-duality transformation on the $x_3$ axis, then smeared twist along $x_4, ..., x_9$ and finally  apply the T-duality on the $x_3$ axis.   In the large $N$ limit the geometry is described by \footnote{A simple way to derive the metric (2.2) is that we first rewrite the part of metric of  (2.1) as $ h(r)^{-1}dr^2+ r^2 d\Omega_5^2 = (h(r)^{-1}-1) dr^2+ (dr^2 + r^2 d\Omega_5^2)  = (h(r)^{-1}-1) dr^2+ (dx_4^2+...++x_9^2)$.  As the smeared twist could only add some terms propositional to $d\theta_i$ it does not change the value of $dr$.  Thus, after the twist along $x_4, ..., x_9$ we could combine the $(h(r)^{-1}-1) dr^2$ with $dr^2$ to the term $\left(1-{U_T^4\over U^4}\right)^{-1}dU^2$ shown in (2.2).}
$$ds_{10}^2 = U^2\left[- \left( 1-{U_T^4\over U^4}\right)dt^2+ dx^2+ dy^2+{ dz^2\over 1+B^2U^2\sin^2\theta_1\sin^2\theta_2}\right]\hspace{4cm}$$
$$+ {1\over U^2} \left[\left( 1-{U_T^4\over U^4}\right)^{-1}dU^2+ U^2d\Omega_5^2-U^4B^2\sin^4\theta_1\sin^4\theta_2 {(a_3d\theta_3+a_4d\theta_4+a_5d\theta_5)^2\over 1+U^2B^2\sin^2\theta_1\sin^2\theta_2}\right]. \eqno{(2.2)}$$
$$e^{2\Phi}= {1 \over  1+ U^2B^2\sin^2\theta_1\sin^2\theta_2},~~~
B_{z\theta_i}d\theta_i = - {U^2B\sin^2\theta_1\sin^2\theta_2 \over 1+U^2B^2\sin^2\theta_1\sin^2\theta_2 }~a_id\theta_i,\hspace{1.7cm}\eqno{(2.3)}$$
in which $a_3 \equiv \cos\theta_4 $, $a_4 \equiv - \sin\theta_3\cos\theta_3\sin\theta_4 $, and $a_5 \equiv \sin^2\theta_3\sin^2\theta_4$, where $\theta_i$ are the angular coordinates parameterizing the sphere $S^5$ transverse to the D3 brane.  Thus there is a nonzero B field with one leg along the brane worldvolume and others transverse to it.  The value $B$ in (2) is proportional to the dipole length $\ell$ defined in the ``non-commutative dipole product" : $\Phi_a (x) * \Phi_a (x) = \Phi_a (x-\ell_b/2) ~\Phi_b (x +\ell_a/2) $ for the dipole field $\Phi(x)$ [18]. 

\subsection {Temporal Wilson Loop}
  To investigate the Wilson loop on the  finite temperature non-commutative dipole field theory  in the dual string description we parameterize the string configuration by
$$\tau=t,~~~~~~~U=\sigma,~~~~~~~z=z(\sigma),\eqno{(2.4)}$$
the Nambu-Goto action becomes
$$S= {1\over 2\pi}\int d\sigma d\tau \left(\sqrt{- det g} +B_{\mu\nu}\partial_\tau X^{\mu}\partial_\sigma X^{\nu}\right)={T_0\over 2\pi}\int d\sigma \sqrt{1+{(U^4-U_T^4) (\partial_\sigma z)^2\over 1+ B^2U^2}},\eqno{(2.5)}$$
in which $T_0$ denotes the time interval we are considering and we have set $\alpha'=1$.   In above calculation we have let $\theta_1=\theta_2 = \pi/2$.  Note that the second term in Nambu-Goto action (2.5) does not contribute in the static case while it will play an important role in the rotating case which is investigated  in section V.  

 As the associated Lagrangian $({\cal L})$ does not explicitly depend on $z$ the function ${\partial{\cal L}\over \partial(\partial_\sigma z)}$ will be proportional to an integration constant, i.e.
$${\partial{\cal L}\over \partial(\partial_\sigma z)} ={{(U^4-U_T^4) (\partial_\sigma z)\over 1+ B^2U^2} \over \sqrt{1+{(U^4-U_T^4) (\partial_\sigma z)^2\over 1+ B^2U^2}}} = {\sqrt{U_0^4-U_T^4} \over\sqrt{1+B^2U_0^2}},\eqno{(2.6)}$$
as at $U=U_0$ we have the property of  $(\partial_\sigma z) \rightarrow \infty$. From above relation we can find the function $(\partial_\sigma z)^2$
$$(\partial_\sigma z)^2 ={{1+B^2U^2\over U^4-U_T^4 }\over {U^4-U_T^4 \over U_0^4 -U_T^4}~{1+B^2U_0^2 \over 1+B^2U^2}-1} .\eqno{(2.7)}$$
Now, we put a quark at place $z=\sigma =-L/2$ and an anti-quark at $z=\sigma = L/2$,  thus
$$L = 2 \int_0^{L/2} d z = 2 \int_{U_0}^\infty dU (\partial_\sigma z)={2\over U_0}\int_1^\infty dy {\sqrt{{1+y^2B^2U_0^2\over y^4 - \left(U_T^4/U_0^4\right)}}\over \sqrt{{y^4 - \left(U_T^4/U_0^4\right)\over 1- \left(U_T^4/U_0^4\right)}~{1+B^2U_0^2\over 1+B^2U_0^2 y^2}-1}}.\eqno{(2.8)}$$
Above relation implies that
$$L \approx  \{\begin{array} {cc}
0& as~U_0\rightarrow~U_T ,\\
B \pi - {\pi \over 4B U_0^2}& as~U_0\rightarrow~\infty. 
\end{array}\eqno{(2.9)}$$
Thus the interquark distant $L$ will asymptotically  approach to a constant $L_0\equiv B \pi$ as $U_0\rightarrow \infty$.   This indicates that it exists a minimum distance between the quark and anti-quark, contrast to that without dipole in which the interquark distant $L$ could approach to zero [14,15]. 

 Note that on the near extremal D-brane background a string shall end at the horizon, $U_0=U_T$, and not at $U_0=0$ [26].   Thus the minimum value of $U_0$ adopted in (2.9) is at $U_0=U_T$.

We can evaluate the interquark potential $H$ form the Nambu-Goto action (2.5) with a help of (2.7).  The formula is
$$H = {1\over \pi}\left[U_0~\int_1^\infty dy \left(\sqrt{(y^4 - \left(U_T^4/U_0^4\right))(1+B^2U_0^2)\over (y^4 - \left(U_T^4/U_0^4\right)) (1+B^2U_0^2) - (1- \left(U_T^4/U_0^4\right)) (1+B^2U_0^2 y^2) }-1\right)\right. $$
$$\left.-U_0+U_T\right].~~~~~~~~\eqno{(2.10)}$$
Here we have subtracted the infinity coming from the mass of W-boson which corresponding to the string stretching from $U=U_T$ to $U=\infty$ [14-16].  Above relation implies that
$$H \approx  \{\begin{array} {cc}
0& as~U_0\rightarrow~U_T ,\\
{U_T \over \pi} - {1 \over 4\pi B U_0}& as~U_0\rightarrow~\infty. 
\end{array}\eqno{(2.11)}$$
Thus the interquark potential $H$ will asymptotically  approach to a constant ${U_T\over  \pi}$ as $U_0\rightarrow \infty$, contrast to that without dipole in which  $H$ will asymptotically  approach to $- \infty$.    The appearances of terms ${\pi \over 4 B U_0^2}$ in (2.9) and ${1 \over 4\pi B U_0}$ in (2.11) reveals the fact of non-perturbative behavior at $B \rightarrow 0$.  Thus, there shall have a qualitative difference between the theories with and without dipole. 

For a clear illustration we first show in figure 1 the functions $L(U_0)$, $H(U_0)$  and $H(L)$ at $(B, U_T) =(0,1)$, which are obtained by performing the numerical evaluation of (2.8) and (2.10). 

\hspace{1cm}

\scalebox{1}{\hspace{2cm}\includegraphics{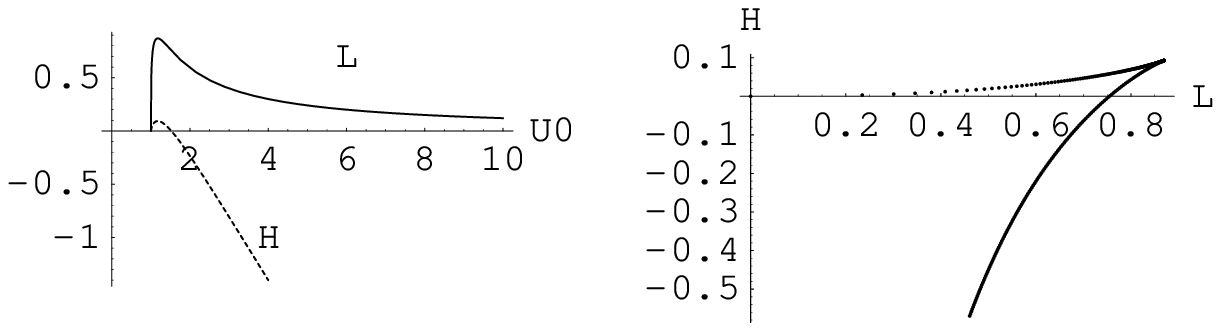}}
\\
\\
{\it Figure 1.  The functions $L(U_0)$, $H(U_0)$  and $H(L)$ at $(B, U_T) =(0,1)$.  There exists a maximum distance and beyond which the  the quark pair will become free due to screening by thermal bath.  The local extremity on $L(U_0)$ and $H(U_0)$ corresponds to the spike on diagram $H(L)$.}

\hspace{1cm}

Let us make following comments about figure 1.
\\
1. Figure 1 is just the case of $B=0$ which was studied in [14-15] while plotted in a coordinate different from them.
\\
2. From figure 1 we see that increasing the turn point $U_0$ from $U_T$ the interquark distance $L$ will be increasing and in this case the interquark potential is also a positive increasing function.   However, after the distance reaches its maximum value it will turn to a decreasing function of $U_0$.   In this situation the positive interquark potential is a decreasing function.  Finally, the interquark potential becomes negative.  Thus, the dual string with two different point value of $U_0$  may correspond to the same interquark distance $L$ which, however, have different interquark potential.  The dual string configuration which has a small potential is more stable and dual to the physical quark system.
\\
3. Figure 1 shows that there exists a maximum distance $L_{max}\approx 0.8$ and we encounter two regions with different behavior.  For $L < 0.7 $ we observe a Coulomb like behavior.   However, when  $  0.7 <L < 0.8$ the dual  string configuration has a positive energy which, however, does not correspond to the lowest energy configuration.   In this case it is energetically favorable for the system to be in a configuration of two parallel string ending on the horizon, which corresponds to zero energy (after the subtraction).  Thus, the dual quarks have zero energy and become free due to screening by thermal bath.   The property was found in  [14-16]. 
\\
4. It is interesting to see that the dual string does not exist after $L > L_{max}$.  Of course, two parallel string ending on the horizon with displacement $L > L_{max}$ could be formed, which, as we know, has zero energy (after the subtraction).   Thus the quarks will become totally free if the interquark distance is too large.  In this case the physical picture is, as mentioned above, the quarks become free due to screening by thermal bath.
\\
\\

In figure 2  we show the functions $L(U_0)$, $H(U_0)$  and $H(L)$ at $(B, U_T) =(0.2, 1)$.

\hspace{1cm}

\scalebox{1}{\includegraphics{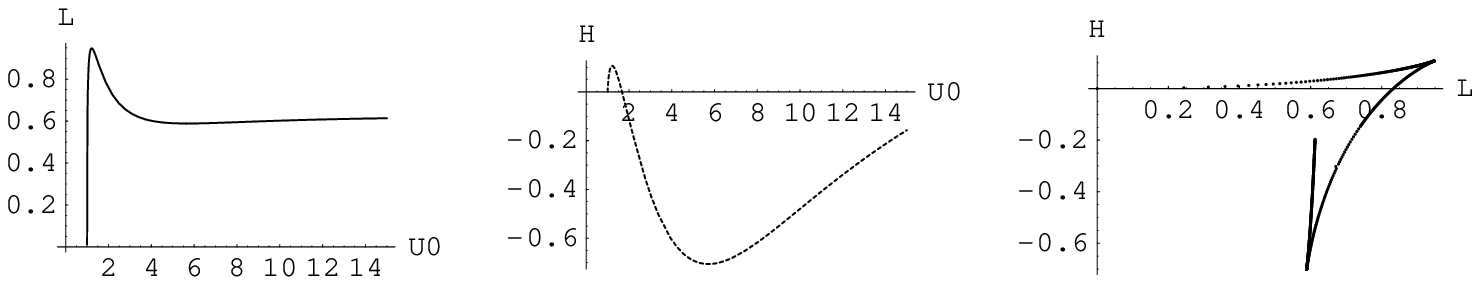}}
\\
\\
{\it Figure 2.  The functions $L(U_0)$, $H(U_0)$  and $H(L)$ at $(B, U_T) =(0.2, 1)$.  There exist a maximum and a minimum distance and the quark boundary pair could be found only if their distance is between the values. The local extremities on $L(U_0)$ and $H(U_0)$ correspond to the spikes on diagram $H(L)$.}

\hspace{1cm}

Figure 2 shows that, like that in the theory without dipole, there exists a maximum distance $L_{max}$ and beyond which the quark boundary pair will become free due to screening by thermal bath, as that in the non-dipole field system.  However, in the presence of dipole field there will exist a minimum distance $L_{min}$, which is an increasing function of dipole field $B$.  Thus, the quark boundary pair could be formed only if their distance is between $L_{min}$ and $L_{max}$.  
\\

Note that in the dipole theory there will exist a minimum distance $L_{min}$.  However, as the maximum distance $L_{max}$ is a decreasing function of temperature the value of  $L_{min}$ may be coincident with $L_{max}$ at sufficiently high temperature.  This means that in the dipole theory the quark boundary pair will become totally free due to screening by thermal bath at sufficiently high temperature.  In this situation we could not find any quark boundary state, in contrast to the non-dipole theory in which the quark pair could be formed in the Coulomb phase at short distance, as shown in figure 1.   

\section {Spatial Wilson Loop in Finite-Temperature Dipole Theory }
The  non-supersymmetric dipole theories at zero temperature could be investigated by considering  the spatial Wilson loop in the background 
of Euclidean near-extremal Dp-brane solutions [13,16].  This is because that when the spatial size is much larger then $1/T$ ($T$ is the Hawking temperature of the near-extremal solution and Euclidean time is compactified along a circular of radius $1/2\pi T$.)  the effective low energy theory reduces effectively to a p-dimensional non-supersymmetric theory. Therefore, the spatial Wilson loop gives us the energy between a quark and an anti-quark of the p-dimensional non-supersymmetric theory at zero temperature.

  We first use metric (2.2) to study 3D non-supersymmetric theory in 3.1 and then construct 4d metric to study 4D non-supersymmetric theory in 3.2.
\subsection {3D Non-supersymmetric Theory}
To investigate the spatial Wilson loop  we parameterize the string configuration by
$$\tau=x~(or~y) ,~~~~~U=\sigma,~~~~~~~z=z(\sigma),\eqno{(3.1)}$$
the Nambu-Goto action calculated  form the metric (2.2) becomes
$$S={T_0\over 2\pi}\int d\sigma \sqrt{{1 \over1- \left(U_T^4/U_0^4\right)}+{U^4 (\partial_\sigma z)^2\over 1+ B^2U^2}}.\eqno{(3.2)}$$
As the associated Lagrangian $({\cal L})$ does not explicitly depend on $z$ the function ${\partial{\cal L}\over \partial(\partial_\sigma z)}$ will be proportional to an integration constant, i.e.
$${\partial{\cal L}\over \partial(\partial_\sigma z)} ={{U^4 (\partial_\sigma z)\over 1+ B^2U^2} \over \sqrt{{1 \over1- \left(U_T^4/U_0^4\right)}+{U^4 (\partial_\sigma z)^2\over 1+ B^2U^2}}} = {U_0^2 \over\sqrt{1+B^2U_0^2}},\eqno{(3.3)}$$
as at $U=U_0$ we have the property of  $(\partial_\sigma z) \rightarrow \infty$. From above relation we can find the function $(\partial_\sigma z)^2$
$$(\partial_\sigma z)^2 ={{1+B^2U^2\over U^4-U_T^4 }\over {U^4 \over U_0^4}~{1+B^2U^2 \over 1+B^2U_0^2}-1} .\eqno{(3.4)}$$
Put a quark at place $z=\sigma =-L/2$ and an anti-quark at $z=\sigma = L/2$,  then
$$L = 2 \int_{U_0}^\infty dU (\partial_\sigma z)={2\over U_0 \sqrt{1+B^2 U_0^2 }}\int_1^\infty dy {{\sqrt{y^4\over y^4 - \left(U_T^4/U_0^4\right)}}\over \sqrt{\left({y^4 \over 1+B^2U_0^2 y^2}\right)^2 - {y^4 \over 1+B^2U_0^2 y^2} {1 \over 1+B^2U_0^2}}}.\eqno{(3.5)}$$
The interquark potential $H$ calculated form the Nambu-Goto action (2.5) with a help of (3.4) becomes 
$$H = {1\over \pi}\left[U_0~\int_1^\infty dy \left( {\sqrt{{y^4\over y^4 - \left(U_T^4/U_0^4\right)}{y^4\over 1+ B^2 U_0^2 y^2}}\over \sqrt{{y^4 \over 1+B^2U_0^2 y^2} - {1 \over 1+B^2U_0^2}}}-1\right) - U_0 + U_T\right]. \eqno{(3.6)}$$

Equation (3.5) implies  that
$$L \approx  \{\begin{array} {cc}
\infty& as~U_0\rightarrow~U_T ,\\
B \pi - {\pi \over 4B U_0^2}& as~U_0\rightarrow~\infty. 
\end{array}\eqno{(3.7)}$$
Thus the interquark distant $L$ will asymptotically  approach to a constant $L_0\equiv B \pi$ as $U_0\rightarrow \infty$.   This indicates that it exists a minimum distance between the quark and anti-quark, contrast to that without dipole in which interquark distant $L$ could approach zero [16]. 

Equation (3.6) implies  that
$$H \approx  \{\begin{array} {cc}
\infty& as~U_0\rightarrow~U_T ,\\
{U_T \over \pi} - {1 \over 4\pi B U_0}& as~U_0\rightarrow~\infty. 
\end{array}\eqno{(3.8)}$$
\\
Thus the interquark potential $H$ will asymptotically  approach to a constant ${U_T\over  \pi}$ as $U_0\rightarrow \infty$, contrast to that without dipole in which  $H$ will asymptotically  approach to $- \infty$.   For a clear illustration we first show in figure 3 the functions $L(U_0)$, $H(U_0)$  and $H(L)$ at $(B, U_T) =(0,1)$. 

\hspace{1cm}

\scalebox{1}{\hspace{1cm}\includegraphics{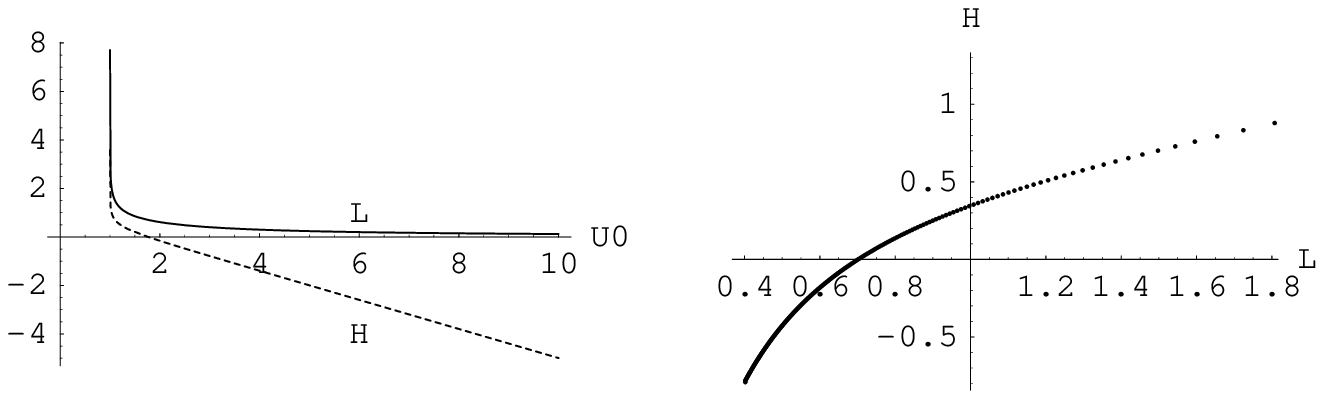}}
\\
\\
{\it Figure 3.  The functions $L(U_0)$, $H(U_0)$  and $H(L)$ at $(B, U_T) =(0,1)$. The linear potential is shown at large distance $L$.}

\hspace{1cm}

Figure 3 shows the linear potential at large distance $L$.  The quark pair could   be formed at any distance.

In figure 4  we show the functions $L(U_0)$, $H(U_0)$  and $H(L)$ at $(B, U_T) =(0.4, 1)$.

\hspace{1cm}

\scalebox{1}{\hspace{2cm}\includegraphics{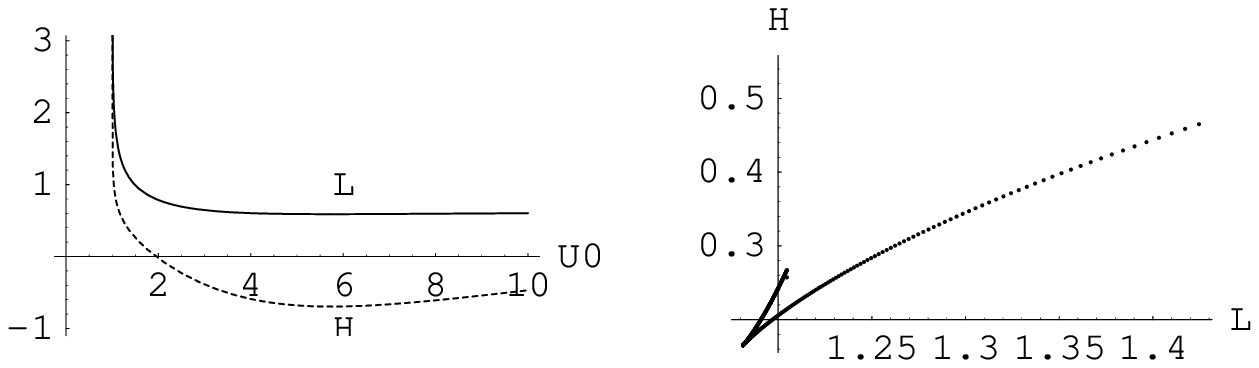}}
\\
\\
{\it Figure 4.  The functions $L(U_0)$, $H(U_0)$  and $H(L)$ at $(B, U_T) =(0.4, 1)$. While the linear potential is shown at large distance $L$ there exist a minimum distance between quarks.  The local extremity on $L(U_0)$ and $H(U_0)$ corresponds to the spike on diagram $H(L)$.}

\hspace{1cm}

Figure 4 shows the linear potential at large distance $L$.  However, the quark pair could  be formed only at distance $L > L_{min}$. 

Note that,  in the limit $U_0 \rightarrow U_T$ both of $L$ and $H$ will approach to $\infty$ and we have the following simple relations 
$$L \rightarrow {2 \sqrt{1+B^2 U_T^2}\over U_T }\int_1^\infty {dy \over y^4 -1}.\eqno{(3.9)}$$
$$H \rightarrow { U_T\over \pi }\int_1^\infty {dy \over y^4 -1}.~~~~~~~~~~~~~ \eqno{(3.10)}$$
Thus we find a linear confined potential at large distance
$$ H = {U_T^2\over 2\pi \sqrt{1+B^2 U_T^2 }} L= { \pi T^2 \over 2 \sqrt{1+B^2 \pi^2 T^2 }}~ L,\eqno{(3.11)}$$
in which we have used the relation $ U_T = \pi T$.   The tension of the QCD string is 
$$ \sigma = { \pi T^2 \over 2 \sqrt{1+B^2 \pi^2 T^2 }},\eqno{(3.12)}$$
which is a decreasing function of dipole field $B$.  This means that although in the limit $LT \gg 1$ the nature of confinement could be shown in the theory with and without dipole, the dipole field will decrease the string tensor. 

\subsection {4D Non-supersymmetric Theory}
To consider the  non-supersymmetric 4D dipole theory at zero temperature we shall consider the supergravity background which is constructed form near-extremal D4-brane solutions, instead of D3-brane.  We could follow the prescription of 2.1 to find the proper background which is described by 
$$ds_{10}^2 = U^{3/2}\left[- \left( 1-{U_T^3\over U^3}\right)dt^2+ dw^2+ dx^2+ dy^2+{ dz^2\over 1+B^2U^2\sin^2\theta_1}\right]\hspace{4cm}$$
$$+ {1\over U^{3/2}} \left[\left( 1-{U_T^3\over U^3}\right)^{-1}dU^2+ U^2d\Omega_4^2-U^4B^2\sin^4\theta_1 {(a_2d\theta_2+a_3d\theta_3+a_4d\theta_4)^2\over 1+U^2B^2\sin^2\theta_1}\right]. \eqno{(3.13)}$$
$$e^{2\Phi}= {U^{3/2} \over  1+ U^2B^2\sin^2\theta_1},~~~
B_{z\theta_i}= - {a_i~U^2B\sin^4\theta_1\over 1+U^2B^2\sin^2\theta_1},\hspace{1.7cm}\eqno{(3.14)}$$
in which $a_2 \equiv \cos\theta_3 $, $a_3 \equiv - \sin\theta_2\cos\theta_2\sin\theta_3 $, and $a_4 \equiv \sin^2\theta_2\sin^2\theta_3$, where $\theta_i$ are the angular coordinates parameterizing the sphere $S^4$ transverse to the D4 brane.  

To investigate the spatial Wilson loop  we parameterize the string configuration by
$$\tau=w~(or~x,~y) ,~~~~~U=\sigma,~~~~~~~z=z(\sigma),\eqno{(3.15)}$$
the Nambu-Goto action calculated form the metric (3.13) becomes
$$S={T_0\over 2\pi}\int d\sigma \sqrt{{1 \over1- \left(U_T^3/U_0^3\right)}+{U^3 (\partial_\sigma z)^2\over 1+ B^2U^2}},\eqno{(3.16)}$$
and the function $(\partial_\sigma z)^2$ calculated as before becomes 
$$(\partial_\sigma z)^2 ={{1+B^2U^2\over U^3-U_T^3 }\over {U^3 \over U_0^3}~{1+B^2U^2 \over 1+B^2U_0^2}-1} .\eqno{(3.17)}$$
The interquark distance is 
$$L = 2 \int_{U_0}^\infty dU (\partial_\sigma z)={2\over U_0 \sqrt{1+B^2 U_0^2 }}\int_1^\infty dy {{\sqrt{y^3\over y^3 - \left(U_T^3/U_0^3\right)}}\over \sqrt{\left({y^3 \over 1+B^2U_0^2 y^2}\right)^2 - {y^3 \over 1+B^2U_0^2 y^2} {1 \over 1+B^2U_0^2}}}.\eqno{(3.18)}$$
The interquark potential $H$ calculated form the Nambu-Goto action (2.5) with a help of (3.17) becomes 
$$H = {1\over \pi}\left[U_0~\int_1^\infty dy \left( {\sqrt{{y^3\over y^3 - \left(U_T^3/U_0^3\right)}{y^3\over 1+ B^2 U_0^2 y^2}}\over \sqrt{{y^3 \over 1+B^2U_0^2 y^2} - {1 \over 1+B^2U_0^2}}}-1\right) - U_0 + U_T\right]. \eqno{(3.19)}$$
Using (3.18) and (3.19) we could plot the  diagrams of  $L(U_0)$, $H(U_0)$  and $H(L)$ which are qualitatively like Figure 4.  Thus it will show the linear potential at large distance $L$ and the quark pair could  be formed only at distance $L > L_{min}$. 

As that in 3D theory,  in the limit $U_0 \rightarrow U_T$ both of $L$ and $H$ will approach to $\infty$ and we have the following simple relations 
$$L \rightarrow {2 \sqrt{1+B^2 U_T^2}\over \sqrt{U_T} }\int_1^\infty {dy \over y^3 -1}.\eqno{(3.20)}$$
$$H \rightarrow { U_T\over \pi }\int_1^\infty {dy \over y^3 -1}.~~~~~~~~~~~~~ \eqno{(3.21)}$$
Thus we find a linear confined potential at large distance
$$ H = {U_T^{3/2}\over 2\pi \sqrt{1+B^2 U_T^2 }} L= { \left(4\pi T/3\right)^3 \over 2\pi  \sqrt{1+B^2 \left(4\pi T/3\right)^4 }}~ L,\eqno{(3.22)}$$
in which we have use the relation $ U_T = \left(4\pi T/3\right)^2$, (which is obtained from the relation $T \equiv {1\over 4\pi} {dg_{tt}\over dU}|_{U=U_T} $).   The tension of the QCD string is therefore 
$$ \sigma = { \left(4\pi T/3\right)^3 \over 2\pi  \sqrt{1+B^2 \left(4\pi T/3\right)^4 }},\eqno{(3.23)}$$
which is a decreasing function of dipole field $B$.  

In conclusion, we have studied the spatial Wilson loop in the high temperature limit of D3-brane and D4-brane background and find the confining nature in the zero-temperature 3D and 4D non-supersymmetry dipole gauge theory.  The string tension of the linear confinement potential we obtained is found to be a decreasing function of the dipole field.  

\section {t'Hooft Loop in 4D Non-supersymmetric Dipole Theory}
In this section we follow the method in [16] to  investigate the associated t'Hooft  loop in the 4D non-supersymmetry dipole theory which shows the nature of quark confinement as proved in (3.21).

  The string theory realized of the monopole  is the D2-brane ending on the D4-brane.  For the metric (3.13) the D2-brane is wrapped along $x_0$ and action  becomes
$$S= {1\over (2\pi)^{3/2}}\int d\sigma_1 d\sigma_2 d\tau ~e^{-\Phi}~\sqrt{- det g} ={T_0~\sqrt{1+ B^2U^2}\over (2\pi)^{3/2}}\int d\sigma \sqrt{1 + \left(U^3 -U_T^3\right){ (\partial_\sigma z)^2\over 1+ B^2U^2}},\eqno{(4.1)}$$
and the function $(\partial_\sigma z)^2$ calculated as before becomes 
$$(\partial_\sigma z)^2 ={{1+B^2U^2\over U^3-U_T^3 }\over {U^3-U_T^3 \over U_0^3-U_T^3}-1} .\eqno{(4.2)}$$
The interquark distance is 
$$L = 2 \int_{U_0}^\infty dU (\partial_\sigma z)={2 \sqrt{1-(U_T^3/U_0^3)}\over  \sqrt{U_0}}\int_1^\infty dy {{\sqrt{1+ B^2U_0^2 y^2\over y^3 - \left(U_T^3/U_0^3\right)}\over \sqrt{y^3 - 1}}}.\eqno{(4.3)}$$
The interquark potential $H$ calculated form the action (4.1) with a help of (4.2) becomes 
$$H = {2\over (2\pi)^{3/2}}\left[U_0~\int_1^\infty dy  \sqrt{1+B^2 U^2}\left( \sqrt{y^3- \left(U_T^3/U_0^3\right)\over y^3-1}-1\right) + \int_{U_0}^{U_T} dU \sqrt{1+B^2 U^2}\right]. \eqno{(4.4)}$$
\\

Let us first consider the case of $ B=0$.  In this case Eq.(4.3) implies that 
$$L \approx  \{\begin{array} {cc}
0& as~U_0\rightarrow~U_T ,\\
0& as~U_0\rightarrow~\infty. 
\end{array}\eqno{(4.5)}$$
Thus  there will exist a maximum distance between the monopole anti-monopole pair configuration, as shown in figure 5.  As the monopole anti-monopole pair configuration could only exist if their distance is less then the minimum distance in the case of non-dipole theory,  in the region  $LU_T \gg 1$ the system would become the free monopole anti-monopole pair configuration.   This is the trivial configuration of two parallel D2-branes ending on the horizon and wrapping along $x_0$.  The screening of magnetic charge thus indicates the confinement of the electric charge and the quark confinement. The property was first shown in [16].  Note that as our arguments are based on the existence of  a maximum distance, contrast to the positive energy result used in [16], it is more easy to see the screening property.   

\hspace{1cm}

\scalebox{1}{\hspace{6cm}\includegraphics{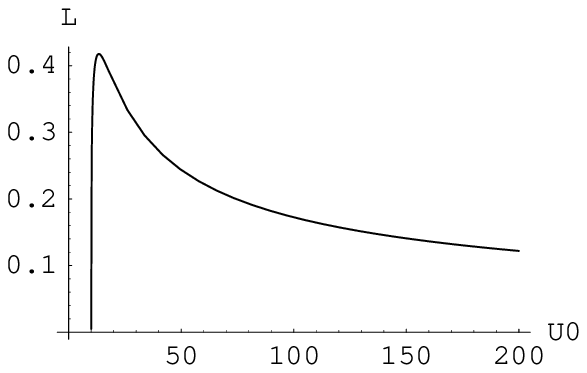}}
\\
\\
{\it Figure 5.  The functions $L(U_0)$ for the case of $(B, U_T) =(0,10)$.  Note that there exist a minimum distance between the monopole anti-monopole pair configuration.}

\hspace{1cm}

\hspace{1cm}

In the case of dipole theory Eqs.(4.3) and (4.4) imply that 
$$\begin{array} {cc}
L \rightarrow { B \sqrt {\pi U_0}\Gamma[1/6] \over  \Gamma[5/3]}& as~U_0\rightarrow~\infty,\\ \\
H \rightarrow - {B U_0^2 \over (2\pi)^{3/2} }~~~~& as~U_0\rightarrow~\infty. 
\end{array}\eqno{(4.6)}$$
Thus there is a repulsive force between the monopole and anti-monopole.   In figure 6 we plot  a diagram of  $L(U_0)$, $H(U_0)$  and $H(L)$ for the case of $(B, U_T) =(0.2,10)$.

\hspace{1cm}

\scalebox{1}{\hspace{.5cm}\includegraphics{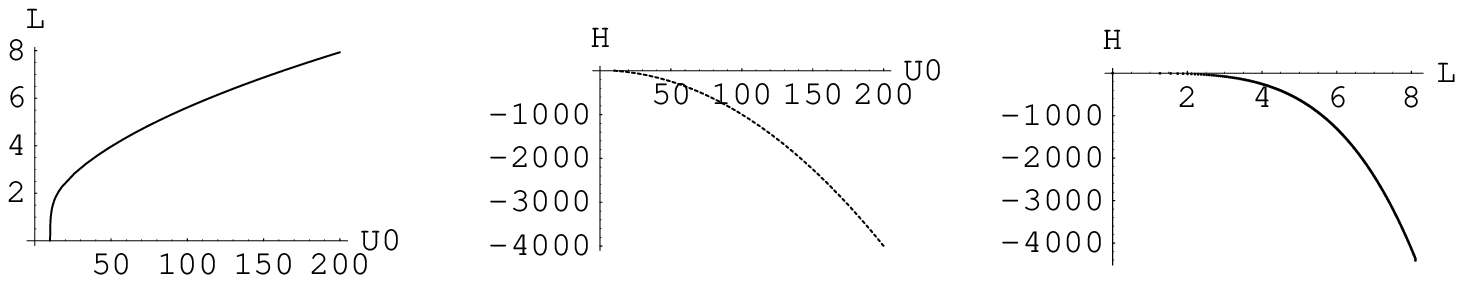}}
\\
\\
{\it Figure 6.  The functions $L(U_0)$, $H(U_0)$  and $H(L)$ at $(B, U_T) =(0.2,10)$.  The monopole and anti-monopole shows strong repulsive force at $LU_T \gg1$.}

\hspace{1cm}

We see that, in the case of non-supersymmetric dipole theory the monopole and anti-monopole shows strong repulsive force at $LU_T \gg1$. 

Let us make following comments to conclude this section.
\\
1.  The Wilson loop studied in 3.2 shows that, for the  4D non-supersymmetric  non-dipole theory, the quarks are in confinement phase which is extended to all distance (shown in figure 3).   The t'Hooft loop studied in [16] and this section shows that, for the  4D non-supersymmetric non-dipole theory, the monopoles are in free phase which have a finite maximum distance (shown in figure 5). 
\\
2.  The Wilson loop studied in 3.2 shows that, for the  4D non-supersymmetric  dipole theory, the quarks are in confinement phase which could be found only if their distance is larger than a critical value (shown in figure 4).   The t'Hooft Loop studied in this section shows that, for the 4D non-supersymmetric dipole theory, the monopoles are in strong repulsive force phase which is extended to all distance  (shown in figure 6). 
\\
3. It is important to mention that, as discussed in [16],  we can trust the supergravity description only if  $TL < N^{2/3}$ ($N$ is the number of D4-brane which is used to construct the metric of (3.13)).  Thus, although  the behavior of monopole pair shows a repulsive force for long distance it is useful only at finite $L$.  This means that it dose not have to show a divergent in the energy.  The property at large distance, however, could not be understood form the dual supergravity.   The problem needs furthermore investigations. 
 
\section {Rotating String Configurations}
As the NS-NS B field appears in the background we shall consider the effect of it on the dual string.   In this situation we shall turn to investigate the dual string configuration which is moving with an angular velocity $\omega$ along the angular $\theta_3$.  
\subsection {Temporal Wilson Loop}
Let us first study the rotating string which corresponds to the static one investigated in 2.2.  We can now parameterize the string configuration by
$$t = \tau,~~~~~~~U=\sigma,~~~~~~~z=z(\sigma),~~~~~~~\theta_3 = \omega \tau,\eqno{(5.1)}$$
the action becomes
$$S= {T_0\over 2\pi}\int d\sigma \sqrt{\left(1 - {\omega^2\over U^2 \left(1-{U^4\over U_T^4}\right)(1+B^2U^2)}\right)\left(1+{(U^4-U_T^4) (\partial_\sigma z)^2\over 1+ B^2U^2}\right)} + {BU^2 \omega ~\partial_\sigma z\over 1+B^2 U^2}.\eqno{(5.2)}$$
As the associated Lagrangian $({\cal L})$ does not explicitly depend on $z$ the function ${\partial{\cal L}\over \partial(\partial_\sigma z)}$ will be proportional to an integration constant, i.e.
$${\partial{\cal L}\over \partial(\partial_\sigma z)} =\sqrt {1 - {\omega^2\over U^2 \left(1-{U^4\over U_T^4}\right)(1+B^2U^2)}}{{(U^4-U_T^4) (\partial_\sigma z)\over 1+ B^2U^2} \over \sqrt{1+{(U^4-U_T^4) (\partial_\sigma z)^2\over 1+ B^2U^2}}}+{BU^2 \omega \over 1+B^2 U^2} = {\omega \over B},\eqno{(5.3)}$$
as at $U \rightarrow \infty $ we have the property of  $(\partial_\sigma z) =0$ and $ U\cdot\partial_\sigma z =0$ to ensure that the end points of the string on the boundary has a finite distance.  From the above relation we can find the function $(\partial_\sigma z)^2$
$$(\partial_\sigma z)^2 ={\omega^2 (1+B^2 U^2)\over B^2 (U^4-U_T^4)}{1\over (1+B^2 U^2)(U^4-U_T^4)- \omega^2 U^2 - {\omega^2\over B^2}}.\eqno{(5.4)}$$
Using the property that $(\partial_\sigma z) \rightarrow \infty $ at $U =U_0 $ we find that
$$ \omega  =  \pm B~\sqrt {U_0^4-U_T^4} .\eqno{(5.5)}$$
Substituting this relation into (5.4) we have a simple relation
$$(\partial_\sigma z)^2 ={U_0^4-U_T^4\over U^4-U_T^4}{1\over U^4-U_0^4}.\eqno{(5.6)}$$
Using (5.5) and (5.6) the action (5.2) could be calculated and the corresponding Hamiltonian is
$$H = \sqrt{{U^4-U_T^4\over U^4-U_0^4}}.\eqno{(5.7)}$$
As (5.6) is just (2.7) and (5.7) is just (2.10) in the case without a dipole field.  Thus the rotating string configuration will correspond to the static case without a dipole field.  

  In conclusion, comparing the quark pair energy function $H(L)$ in figure 1 to that in figure 2 we thus see that, while beyond $L_{max}$ the quark pair is free it will become the boundary state of dipole system as $L< L_{max} $, and below the $L_{min}$ it will transit to the rotating configuration.  As the energy  is discontinuous at $L_{min}$ the transition form the static to the rotating configuration is the first order phase transition.  For clear we present the phase structure of  4D finite-temperature non-commutative dipole theory in the table 1.
\\
\\
{\it Table1.  The phase structure of 4D finite-temperature non-commutative dipole theory.}
\\
\\
\scalebox{1}{\hspace{4.5cm}\includegraphics{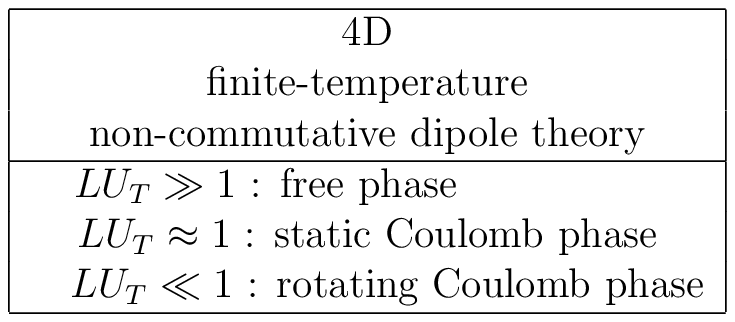}}

\subsection {Spatial Wilson Loop}
Let us next study the rotating string which corresponds to the static one investigated in 3.1.  We can now parameterize the string configuration by
$$\tau=x~(or~y) ,~~~~~U=\sigma,~~~~~~~z=z(\sigma),~~~~~~\theta_3 = \omega \tau,\eqno{(5.8)}$$
the action calculated  form the anastz becomes
$$S= {T_0\over 2\pi}\int d\sigma \sqrt{{1\over \left(1-{U^4\over U_T^4}\right)}\left(1 - {\omega^2\over U^2 (1+B^2U^2)}\right)\left(1+{(U^4-U_T^4) (\partial_\sigma z)^2\over 1+ B^2U^2}\right)} + {BU^2 \omega ~\partial_\sigma z\over 1+B^2 U^2}.\eqno{(5.9)}$$
As before, we can find the function $(\partial_\sigma z)^2$
$$(\partial_\sigma z)^2 ={\omega^2 (1+B^2 U^2)\over B^2 (U^4-U_T^4)}{1\over U^4(1+B^2 U^2)- \omega^2 U^2 - {\omega^2\over B^2}}.\eqno{(5.10)}$$
Using the property that $(\partial_\sigma z) \rightarrow \infty $ at $U =U_0 $ we find that
$$ \omega  =  \pm B~U_0^2.\eqno{(5.11)}$$
It is interesting to see that the above angular velocity $\omega$ in zero-temperature system does not depend on temperature, contrast to that in finite-temperature system shown in (5.5), which depends on temperature $U_T$.

Substituting this relation into (5.10) we have a simple relation
$$(\partial_\sigma z)^2 ={U_0^4\over U^4-U_T^4}{1\over U^4-U_0^4}.\eqno{(5.12)}$$
Using (5.11) and (5.12) the action (5.9) could be calculated and the corresponding Hamiltonian is
$$H = {U^4\over \sqrt{(U^4-U_0^4)(U^4-U_T^4)}}.\eqno{(5.13)}$$
As (5.12) is just (3.4) and (5.13) is just (3.6) in the case without a dipole field.  Thus the rotating string configuration will correspond to the static case without a dipole field.

  Now, from Eq.(3.11) we see that the quark pair of dipole system has lower energy than that without dipole (which corresponds to the rotating configuration), thus at long distance the quarks will be in the confinement phase with dipole field.  However, below the $L_{min}$ it will transit to the rotating configuration.  As the energy is discontinue at $L_{min}$ the transition form the static to the rotating configuration is the first order phase transition.  For clear we present the phase structure of zero-temperature 3D non-supersymmetric  non-commutative dipole theory in the table 2. (Note that the zero-temperature 4D  non-supersymmetric  non-commutative dipole theory has a similar phase structure after the similar analysis.)
\\
\\
{\it Table 2.  The phase structure of zero-temperature non-supersymmetric  non-commutative dipole theory.}
\\\\
\scalebox{1}{\hspace{4.5cm}\includegraphics{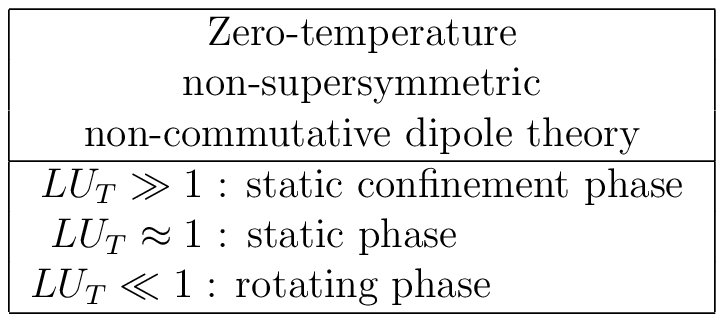}}
\\
\\

Finally, let us make following comments about above result.

1.  It is surprised that the moving string has the same result as that without dipole field.  The reason behind it may be argued  as following.  The dual string  in a background with  $B_{z \theta_3}$ field is somewhat analogous to the situation when a charged particle enters a region with a magnetic field.  Thus, the string will be rotating along $\theta_3$ with a constant angular momentum $\omega$ which is proportional to the strength of the NS-NS B field, as shown in (5.5) and (5.11).   The rotating configuration therefore will have the extra  binding energy (it is negative) from B field which just be canceled by the kinetic energy (it is positive) from the moving.  Thus the rotating dual string does not depend on the  value of dipole field and we have the same result as that without dipole field.  

2.  As the angular velocity of the rotating string configuration is along $\theta_3$ it does not show the angular momentum in the real world spacetime ($(t,x,y,z)$). Thus, the system associated to the rotating string may have intrinsic dynamic (something likes isospin) arising from the dipole.  Our investigations have shown that the rotating dual string dynamics could dramatically change the string behavior and thus the quark system.

\section{Conclusion}
In this paper, we have investigated Wilson loop in the finite-temperature non-commutative dipole field theory from the  string/gauge correspondence.  We first construct the dual supergravity background of the finite temperature non-commutative dipole theory by considering the near-horizon geometry of near-extremal D3-branes, after applying T-duality and  smeared.  We study the temporal Wilson loop and find that while the temperature produces a maximum distance $L_{max}$ between the quarks the dipole field could produce a minimum distance $L_{min}$. The quark boundary pair therefore could be found only if their distance is between $L_{min}$ and $L_{max}$.  We also show that, beyond a critical temperature the quark pair becomes totally free due to screening by thermal bath.  

We next study the spatial Wilson loop in the corresponding D3- and D4-branes background, which dual to the zero temperature 3D and 4D non-supersymmetry dipole gauge theory.  We find that the interquark potential shows the nature of confinement and the string tension is a decreasing function of the dipole field. We also investigate the associated t'Hooft  loop in the 4D non-supersymmetry gauge theory to see the nature of quark confinement.  As the t'Hooft  loop is the ``electric-magnetic" dual of the Wilson loop it  will describe the monopole anti-monopole potential.  We find that the expectation value of t'Hooft  loop shows strong repulsive force between the monopole and anti-monopole, contrast to the conventional of screening of magnetic charge. 

We finally study the dual string which is rotating along the dipole deformed $S^5$ and see that it will behave as a static one without dipole field.  We find that it has no minimum distance and has larger energy than a static one with dipole field. We compare the energy between the static and rotating configurations and find the phase transition between them.  Our results are plotted in tables 1 and 2 which show clearly the phase structure of the finite temperature non-commutative dipole theory and zero-temperature non-supersymmetric non-commutative dipole theory.

\newpage
{\bf  \Large References}
\begin{enumerate}
\item J.~M. Maldacena, ``The large N limit of superconformal field theories  and supergravity,''  Adv. Theor. Math. Phys.  2  (1998) 231-252  [hep-th/9711200].
\item E.~Witten, ``Anti-de Sitter space and holography,'' Adv.\ Theor.\ Math.\ Phys.\   2 (1998) 253 [hep-th/9802150].
\item S.~S.~Gubser, I.~R.~Klebanov and A.~M.~Polyakov, ``Gauge theory correlators from non-critical string theory,'' Phys.\ Lett.\ B 428 (1998) 105
[hep-th/9802109].
\item J.~M. Maldacena,  ``Wilson loops in large N field theories,''  Phys.   Rev. Lett.  80 (1998) 4859-4862 [hep-th/9803002]; S.-J. Rey and J.-T. Yee,  ``Macroscopic strings as heavy quarks in large  N gauge theory and anti-de Sitter  supergravity,''   Eur. Phys. J.   C22 (2001) 379--394 [hep-th/9803001]; J. Gomis and F. Passerini ``Holographic Wilson Loops," JHEP 0608 (2006) 074 [hep-th/0604007].
\item N. Itzhaki, J. M. Maldacena, J. Sonnenschein, S. Yankielowicz, `` Supergravity and The Large N Limit of Theories With Sixteen Supercharges
,'' Phys.Rev. D58 (1998) 046004 [hep-th/9802042]; O. Aharony, A. Fayyazuddin, J. Maldacena, ``The Large N Limit of   N =2,1  Field Theories from Threebranes in F-theory,'' JHEP 9807 (1998) 013 [hep-th/9806159].
\item L. Girardello, M. Petrini, M. Porrati, A. Zaffaroni, ``The Supergravity Dual of N=1 Super Yang-Mills Theory,'' Nucl.Phys. B569 (2000) 451-469 [hep-th/9909047 ]; J. Polchinski and M. J. Strasslerv,'' The String Dual of a Confining Four-Dimensional Gauge Theory,''  [hep-th/0003136 ]; J. Babington, D. E. Crooks, N. Evans,'' A Stable Supergravity Dual of Non-supersymmetric Glue
,'' Phys.Rev. D67 (2003) 066007 [hep-th/0210068 ]; G.V. Efimov, A.C. Kalloniatis, S.N. Nedelko, ``Confining Properties of the Homogeneous Self-Dual Field and the Effective Potential in SU(2) Yang-Mills Theory,'' Phys.Rev. D59 (1999) 014026 [hep-th/9806165].
\item T. Mateos, J. M. Pons, P. Talavera,  ``Supergravity Dual of Noncommutative N=1 SYM,'' Nucl.Phys. B651 (2003) 291-312 [hep-th/0209150]; E. G. Gimon, L. A. P. Zayas, J. Sonnenschein, M. J. Strassler, ``A Soluble String Theory of Hadrons,'' JHEP 0305 (2003) 039 [hep-th/0212061].
\item R. Casero, C. Nunez, A. Paredes,  ``Towards the String Dual of N=1 Supersymmetric QCD-like Theories,'' Phys.Rev. D73 (2006) 086005 [hep-th/0602027].
\item  I.~R.~Klebanov and E.~Witten, ``Superconformal field theory on threebranes at a Calabi-Yau singularity,'' Nucl.\ Phys.\ B536 (1998) 199 [hep-th/9807080].
\item I.~R.~Klebanov and A.~A.~Tseytlin, ``Gravity duals of supersymmetric
SU(N) $\times$ SU(N+M) gauge theories,'' Nucl.\ Phys.\ B 578 (2000) 123 
[hep-th/0002159].
\item I.~R.~Klebanov and M.~J.~Strassler, ``Supergravity and a confining gauge theory: Duality cascades and $\chi$SB-resolution of naked singularities,'' 
JHEP 0008 (2000) 052 [hep-th/0007191].
\item J.~M.~Maldacena and C.~N\'u\~nez, ``Towards the large N limit of pure N = 1 super Yang Mills,'' Phys.\ Rev.\ Lett.\   86 (2001) 588 [hep-th/0008001].
\item E.~Witten, ``Anti-de Sitter space, thermal phase transition, and confinement in  gauge theories,'' Adv.\ Theor.\ Math.\ Phys.\   2 (1998) 505 [hep-th/9803131].
\item S.-J. Rey, S. Theisen and J.-T. Yee,   ``Wilson-Polyakov Loop at Finite Temperature in Large N Gauge Theory and Anti-de Sitter Supergravity,'' Nucl.Phys. B527 (1998) 171-186 [hep-th/9803135].
\item A. Brandhuber, N. Itzhaki, J. Sonnenschein and S. Yankielowicz,  ``Wilson Loops in the Large N Limit at Finite Temperature,'' Phys.Lett. B434 (1998) 36-40 [hep-th/9803137].
\item  A. Brandhuber, N. Itzhaki, J. Sonnenschein and S. Yankielowicz,  ``Wilson Loops, Confinement, and Phase Transitions in Large N Gauge Theories from Supergravity,'' JHEP 9806 (1998) 001 [hep-th/9803263].
\item H. Boschi-Filho, N. R. F. Braga , C. N. Ferreira, ``Heavy quark potential at finite temperature from gauge/string duality," Phys. Rev D74 (2006) 086001 [hep-th/0607038].
\item A. Bergman and O. J. Ganor,``Dipoles, Twists and Noncommutative Gauge Theory," JHEP 0010 (2000) 018 [hep-th/0008030]; A. Bergman, K. Dasgupta, O. J. Ganor, J. L. Karczmarek, and G. Rajesh,``Nonlocal Field Theories and their Gravity Duals," Phys.Rev. D65 (2002) 066005 [hep-th/0103090]. 
\item  K. Dasgupta and M. M. Sheikh-Jabbari, ``Noncommutative Dipole Field Theories," JHEP 0202 (2002) 002 [hep-th/0112064]. 
\item M. Alishahiha and H. Yavartanoo,``Supergravity Description of the Large N Noncommutative Dipole Field Theories," JHEP 0204 (2002) 031 [hep-th/0202131].
\item Wung-Hong Huang, ``Dual String Description of  Wilson Loop in Non-commutative Gauge Theory,''  Phys. Lett. B647 (2007) 519 [hep-th/0701069 ]; ``On the Supergravity Description of Wilson Loop in Non-commutative Dipole Field Theory,''  ArXiv:0706.2410 [hep-th]. 
\item  A. Connes, M. R. Douglas, and A. Schwarz, ``Noncom-mutative geometry and matrix theory: Compactification on tori", JHEP 02 (1998) 003 [hep-th/9711162]. 
\item  M. Douglas. R. and C. Hull,``D-branes and the non-commutative torus", JHEP 02 (1998) 008 [hep-th/9711165]. 
\item N. Sadooghi and M. Soroush, ``Noncommutative  Dipole QED", Int. J. Mod. Phys. A18 (2003) 97 [hep-th/0206009]. 
\item G.T. Horowitz and A.~Strominger, ``Black strings and P-branes,'' Nucl. Phys. B  360 (1991) 197.
\item J.~M. Maldacena,  ``Probing Near Extremal Black Holes with D-branes", Phys. Rev. D57 (1998) 3736 [hep-th/9705053].  
\end{enumerate}
\end{document}